# Quantum molecular dynamics study of the pressure dependence of the ammonia inversion transition

I.M. Herbauts and D.J. Dunstan,
Physics Department, Queen Mary, University of London,
London E1 4NS, England.

**Abstract:** The mechanism of the shift, broadening and quenching of the ammonia inversion frequency with gas pressure has been a problem of lively interest for over seventy years. A simple quantum model of the ammonia molecule perturbed by collisions with ideal gas molecules displays the essential features of the experimental data for $NH_3$ and for $ND_3$. The model does not display the behaviour expected from theories of quantum localisation such as quantum state diffusion and decoherence. On the other hand, models of perturbed classical oscillators do display similar behaviour to our model. The quenching of the ammonia inversion transition cannot therefore be interpreted as spatial localisation of the wavefunction.

*Introduction:*—Since the early days of microwave spectroscopy, the inversion transition of the ammonia molecule NH$_3$ and ND$_3$ has been extensively studied experimentally and theoretically. Much of the interest lies in the fact that it is the smallest and simplest of the pyramidal and enantiomorphic molecules whose ground and excited energy eigenstates are the quantum superpositions of two different spatial configurations, *and* that it is light enough that the transitions between the energy eigenstates are fast enough to be experimentally accessible. The ammonia molecule has two spatial eigenstates $|L\rangle$ and $|R\rangle$ with the nitrogen atom on one side or the other of the plane of hydrogen atoms, and its energy ground and first excited states $|0\rangle$ and $|1\rangle$ are the symmetric and antisymmetric quantum superpositions of the spatial eigenstates (ignoring rotational and vibrational states).

The ammonia maser is based upon the transition between the energy eigenstates, which may also be described as the Rabi oscillation between the spatial eigenstates. However, the inversion transition is seen only at low gas pressure. As the gas pressure is increased, the transition broadens, shifts to lower frequency and then quenches (the frequency goes to zero). The ammonia molecule appears to undergo spatial localisation as a result of interaction with the environment. This would be of immense theoretical interest. In chemistry and in the classical world generally, enantiomorphic molecules with distinguishable spatial eigenstates $|L\rangle$ and $|R\rangle$ are always found in their spatial eigenstates (classical behaviour) rather than their energy ground states (quantum behaviour).[3] Whilst ammonia is not enantiomorphic, it does appear to show both behaviours, quantum at low pressure and classical at high pressure, if the quenching is considered to be a direct observation of localisation or collapse of the wave-function into a spatial eigenstate. Within the context of the decoherence programme, it has been treated quantitatively in that way.[4]

In this paper, we show that interaction with the environment quenches the inversion transition for what might be described as 'classical' reasons. The broadening, shift and quenching of the Rabi oscillation are simply consequences of impacts and may be described within the framework of an oscillator subject to white noise from the environment. There is no evidence for localisation onto spatial eigenstates.

*Background:*—At low pressures in the gas phase, the transition between the energy eigenstates is observed near 24GHz (0.8 cm$^{-1}$) in NH$_3$.[1] In ND$_3$ [2] the transition is near 1.6 GHz (0.053 cm$^{-1}$). In NH$_3$, broadening is observed at pressures above a few mm of mercury, with a shift to lower frequency, and quenching is complete at about 1.7 bar. In ND$_3$, pressures about 15 times lower yield the same effects, in proportion to the inversion transition frequency.

The first explanation of the shift and broadening of the ammonia inversion transition frequency was given by Anderson[5] in terms of perturbation by the electric dipole-dipole interaction between ammonia molecules. Anderson's discussion was only qualitative, and Margenau investigated the quantum states of two ammonia molecules coupled by their dipole-dipole interaction in more detail.[6] He showed that the interaction leads to the splitting of the transition into a higher frequency component with reduced strength and a lower frequency with increased strength. While this accounts for the initial shift to lower frequency, it fails to account for the quenching of the inversion transition at a higher pressure. More recently, the dipole-dipole interaction model has been treated by a quantum mean-field approximation yielding, apparently, a frequency shift, quenching and spatial localisation at pressures for NH$_3$ and ND$_3$ in good agreement with experiment.[7]

The standard theory of line-broadening by impact is given by Van Vleck and Weisskopf.[8] It predicts a line-shape function

$$f(\nu) = \frac{1}{1+b^{-2}(\nu-\nu_0)^2} + \frac{1}{1+b^{-2}(\nu+\nu_0)^2} \tag{1}$$

where the width *b* is given by $1/2\pi\tau$ for strong impacts occurring at a mean interval of $\tau$, and therefore proportional to the pressure. The theory does not predict any peak shift: $\nu_0$ is a constant, the natural frequency of the oscillator. Anderson developed the theory further and obtained a shift of $\nu_0$ to lower frequency equal to the width *b*.[9] Fano recast the problem of pressure broadening in the Liouville representation and obtained a shift to lower frequency independent of the broadening.[10] Ben-Reuven used the Fano theory to show that the ammonia spectra can be well-fitted with a related expression but with three independent parameters proportional to the pressure. Two of them express the effects of elastic collisions on the width and on the frequency shift, and the third parameter expresses the effect of inelastic collisions.[11]

We are interested in a dynamical theory of the transition and of quenching and localisation. It is important to know if the dipole-dipole interaction of ammonia molecules is crucial to the quenching, or if it merely influences the collision cross-section while impacts are sufficient to account for the quenching. Accordingly, we have set up a molecular dynamics simulation in which the quantum nature of the ammonia molecule is explicitly taken into account.[12] Here we show that the model accounts for the shift, broadening and quenching of the inversions transition purely in terms of perturbation by collision with ambient gas molecules.

*The Ammonia Quantum Molecular Dynamics Model:*—We model the problem in one dimension. The ammonia molecule is modelled by a double-well potential, with the two time-dependent spatial wavefunctions $\Psi_L$ and $\Psi_R$. With a weak coupling between the wells the Hamiltonian in the spatial basis is

$$\mathbf{H} = \begin{pmatrix} \omega_0 & \tfrac{1}{2}\omega_1 \\ \tfrac{1}{2}\omega_1 & \omega_0 \end{pmatrix} \tag{2}$$

Diagonalising, the ground and first excited states of the system are found to be $\Psi_0$ and $\Psi_1$ with a frequency splitting of $\omega_1$. The general state of the system is a superposition, with

$$\Psi = a\Psi_0 + b\Psi_1$$
$$|a|^2 + |b|^2 = 1 \tag{3}$$

Expanding this in the spatial basis set $\Psi_L$ and $\Psi_R$, we have time-varying coefficients,

$$\Psi = \alpha(t)\Psi_L + \beta(t)\Psi_R \tag{4}$$

so that the amplitude of the wave-function beats, or oscillates between the two wells. The squared amplitude $|\alpha(t)|^2 = \alpha^*\alpha$ oscillates at the frequency $\omega_1$ and with a beat amplitude that depends on the initial values of *a* and *b* (from zero for e.g. $a = b = 1/\sqrt{2}$ to a maximum amplitude of unity for e.g. $a = 1, b = 0$). This oscillation is the inversion transition or Rabi oscillation of the molecule.

We model impacts, or interactions with the environment, by a term which is diagonal in the spatial representation. That is, we suppose that the double well is tilted during an impact. If a gas atom coming in from the left raises the energy of the left-hand well, the Hamiltonian during impact is

$$\mathbf{H}' = \begin{pmatrix} \omega_0 + \omega_P & \tfrac{1}{2}\omega_1 \\ \tfrac{1}{2}\omega_1 & \omega_0 \end{pmatrix} \quad (5)$$

Diagonalising and expanding in the spatial basis set as before, we obtain the normalised eigenvectors **u** and **v** of $\mathbf{H}'$. Equations 3 and 4 become

$$\Psi' = a_P \Psi'_0 + b_P \Psi'_1 = \alpha'(t)\Psi_L + \beta'(t)\Psi_R \quad (6)$$

The Rabi oscillation is now at a much higher frequency and a much smaller amplitude (for $\omega_P \gg \omega_1$). In reality, the perturbation rises and falls continuously in an impact, but we approximate with a top-hat function, so that $\omega_P$ is switched on at a time $t_0$ and switched off again at $t_1$. At these times, we match the coefficients in the spatial basis, using $\alpha'(t_0) = \alpha(t_0)$ and $\beta'(t_0) = \beta(t_0)$ to solve for $a_P$ and $b_P$ at the onset of the perturbation, and then the new $\alpha(t_1) = \alpha'(t_1)$, $\beta(t_1) = \beta'(t_1)$ to solve for the new $a$ and $b$ at the end of the perturbation. These boundary conditions ensure that the amplitude and phase of the wave-function in each well do not change discontinuously at the beginning and end of the perturbation. The resulting time evolution of $\alpha^*\alpha$ is illustrated in Fig.1.

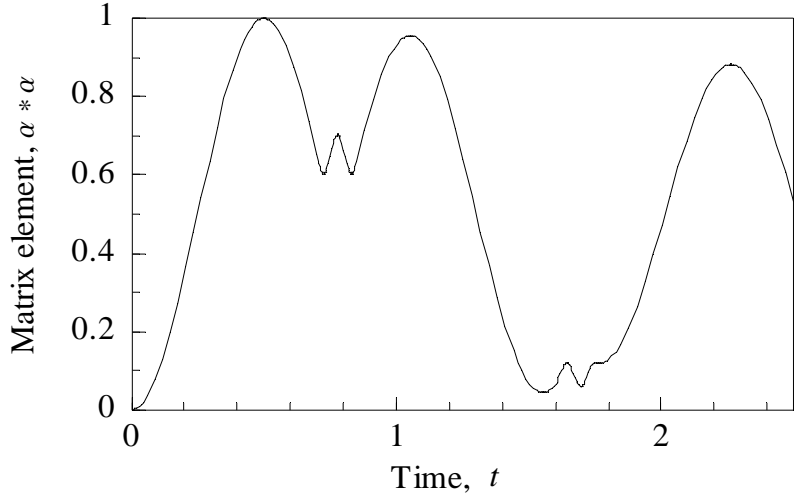

**Fig.1.** The evolution of the occupancy of the left-hand well is shown with two perturbations occurring at $t = 0.7$ and $t = 1.6$. The Rabi angular frequency $\omega_1$ is $2\pi$ and the perturbation $\omega_P = 60$. The initial wavefunction is given by $a = b = 1/\sqrt{2}$; after the two perturbations the values are $a = 0.54 - 0.73i$, $b = 0.36 + 0.22i$.

To model $NH_3$ and $ND_3$, we can choose the units of time so that the Rabi frequency is unity ($\omega_1 = 2\pi$). The strength of the perturbation is of the order of $k_B T$, which at room temperature is 208 cm$^{-1}$. For $NH_3$, therefore, we take $\omega_P = 208\, \omega_1/0.8 = 260\, \omega_1$ and for $ND_3$, $\omega_P = 208\, \omega_1/0.0.053 = 3925\, \omega_1$. The duration $\Delta t = t_0 - t_1$ of an impact is hard to estimate. However, inspection of Fig.1 shows that to achieve a strong impact (in the sense of Van Vleck and Weisskopf [8]), we need something of the order of one cycle of the perturbed Rabi oscillation, i.e. $\omega_P \Delta t \sim 2\pi$, while larger values will have no extra effect. We therefore take values of $\Delta t$ from a random distribution over the range 0 to $2\pi / \omega_P$. The average frequency of impacts corresponds to the gas pressure. We require an impact cross-section to relate the frequency of impacts to the gas pressure quantitatively. Bleaney and Loubser and other authors obtain impact cross-sections from the pressure-broadening of the transition, assuming strong impacts and using $b = 1/2\pi\tau$. We shall see below that such estimates are unreliable, and therefore in our simulation we use the measure $p$ impacts per Rabi cycle instead of pressure, and we vary $p$ over a wide range.

We calculate the values of $\alpha^*\alpha$ at discrete time intervals $\delta t$ with $\Delta t < \delta t \ll 1$. At each time interval we have a probability $\delta t/\tau$ of having an impact, so that there are $p = 1/\tau$ impacts per cycle. If there is an impact, we use $\alpha'(t_0) = \alpha(t_0)$ and $\beta'(t_0) = \beta(t_0)$ to solve for $a_P$ and $b_P$ at the onset of the perturbation, and then calculated the new $\alpha(t_1) = \alpha'(t_1)$, $\beta(t_1) = \beta'(t_1)$ to solve for the new $a$ and $b$ at the end of the perturbation. Then the calculation of the list of values is resumed. Examples are shown in Fig.2 for medium (a) and high (b) values of $p$. The numerical Fourier Transforms of the lists are calculated, shown in Fig.2(c) and (d), and fitted with $Af(\nu)$ of eq.1, with $b$, $\nu_0$ and amplitude $A$ as fitting parameters. Our interest here is the fitted values of $b$ and $\nu_0$ as functions of $p$. In Figure 3 these are compared with the experimental data for $NH_3$,[16] with the constant of proportionality between $p$ and pressure (corresponding to the impact cross-section) as a free parameter.

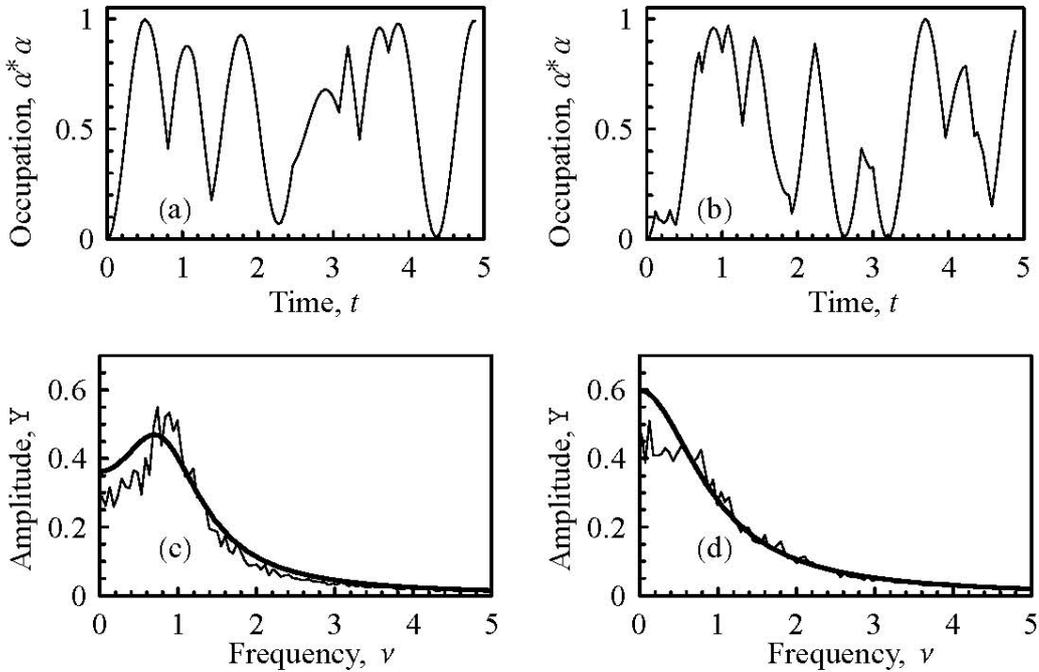

**Fig.2.** The occupation of the left-hand well, $y(t) = \alpha^*\alpha$, is plotted against time, for (a) $p = 3.5$ impacts per cycle, below the quenching, and (b) $p = 7.5$ impacts per cycle, above the quenching. The Fourier

transforms $Y(\nu)$ are shown in (c) and (d) respectively, together with the fits using eq.1.

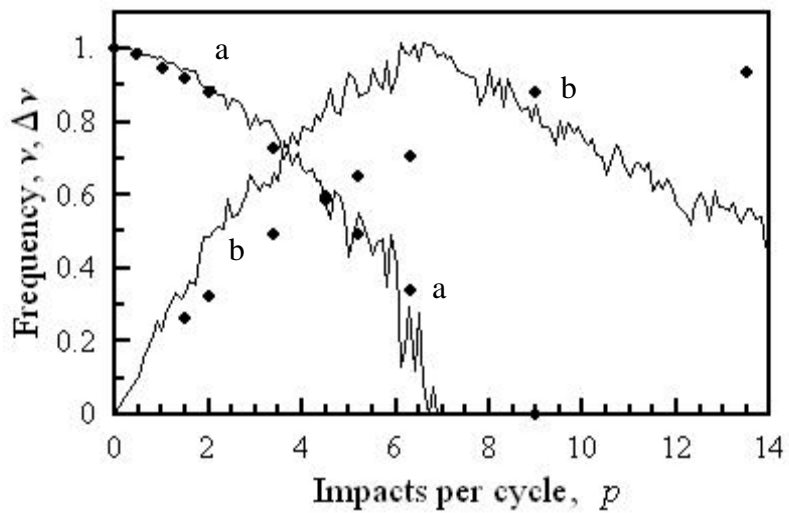

**Fig.3.** The solid curves show (a) the peak frequency and (b) the broadening for $NH_3$ as a function of the number of impacts per cycle as described in the text. The data points show (a) the peak frequency and (b) the broadening reported by Bleaney and Loubser,[16] scaled as described in the text.

The $NH_3$ data is plotted with $p = 4.5$ impacts per cycle equivalent to a pressure $P = 1$ bar. For $ND_3$, the data fits equally well but with $p = 4.5$ equivalent to the pressure $P = 1/15$ bar, consistent with the fifteen times lower inversion frequency in $ND_3$ but the same impact parameter. In both cases, full quenching is observed at about 6.5 impacts per cycle. The model presented here accounts remarkably well for the shift and quenching of the ammonia inversion transition peak. It accounts less well for the broadening, which occurs initially at the rate $b \sim 0.25\,p$ in the simulation and $0.18\,p$ in the experimental data. The experimental broadening is about three-quarters of the model broadening up to the quenching pressure, and above the quenching pressure the experimental broadening continues to increase while the model broadening decreases. To gain a better understanding of this behaviour, we investigate how a simple classical oscillator behaves under similar perturbations.

*A Classical Perturbed Harmonic Oscillator:*—A classical oscillator may be perturbed by collision in a large variety of well-defined ways. We evaluate two perturbations here. We calculate the values of a sinusoid of frequency $\nu = 1$ at discrete time intervals $\delta t \ll 1$. At each time interval we have a probability $\delta t/\tau$ of having an impact, so that there are $p = 1/\tau$ impacts per cycle. If there is an impact, the sinusoid is modified accordingly, and then the calculation of the list of values is resumed. The numerical Fourier Transform is calculated and we find that $af(\nu)$ of eq.1 fits well for a variety of definitions of the impacts, over a very wide range of $p$, with $b$, $\nu_0$ and $a$ as fitting parameters. Our interest here is the fitted values of $b$ and $\nu_0$ as functions of $p$.

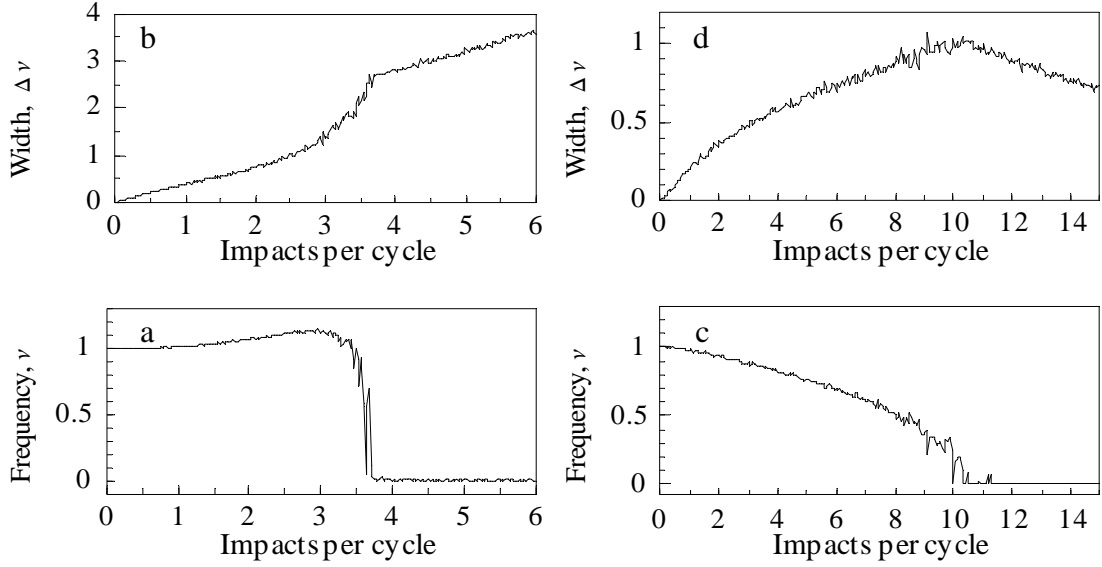

**Fig.4.** The peak frequency and the peak width are plotted for the broken sinusoids with unity frequency described in the text. In (a) and (b), results are for the strongest possible impact, with phase and amplitude completely randomised by the impact. In (c) and (d), the boundary condition at impact keeps the sinusoid continuous but changes the phase and amplitude at random within that constraint.

The strongest perturbation possible (in the sense of Van Vleck and Weisskopf [8]) is a collision that destroys all memory of position and speed (or amplitude and phase). To model this, at impact we pick the new amplitude $A$ of the sinusoid $A\cos(2\pi\nu t + \varphi)$ at random from the range 0 to 1 and the new phase $\varphi$ at random from the range 0 to $2\pi$. In this model, the peak is shifted to *higher* frequency, shown in Fig.4(a) until the width of the transition in Fig.4(b) reaches $b = 1$, at which point quenching occurs, i.e. the frequency collapses to zero. The width continues to increase at still higher values of $p$, Fig.4(b). This is a stronger impact than the impacts of the ammonia molecule, for the quenching occurs at $p = 3.5$ impacts per cycle and the initial slope of the broadening is given by $b \sim 0.5\,p$. It is interesting that Van Vleck and Weisskopf give the broadening for strong impacts as $1/2\pi\tau$, equivalent to $0.16\,p$.

In an alternative definition of impact which is in closer accordance with the ammonia model of Section 2.1 and Fig.1, we define the impact at $t_0$ by taking the position $x(t_0)$ as unchanged by the impact, the new amplitude $A$ as random in the range $x(t_0)$ to 1, and the new phase $\varphi$ such that the speed $\dot{x}(t_0)$ is a random variable in the range consistent with the new amplitude. In this model, the peak shift in Fig.4(c) and width behave in very much the same way as the ammonia results of Fig.3, with the initial broadening $b \sim 0.2\, p$. However, the impact is weaker than in the ammonia model, for quenching occurs at $p = 10.5$ impacts per cycle. Above quenching $b$ decreases again.

In these models we may weaken the strength of the impact by letting the new amplitude and phase be given by some amount $\varepsilon$ of the amplitude and phase calculated as above plus $(1 - \varepsilon)$ of the old amplitude and phase. Not surprisingly, the shift and broadening are identical if plotted as functions of the normalised impact rate $\varepsilon p$.

*Results and Discussion:*—Fig.4 shows that the details of the peak shift and the broadening are very sensitive to the exact nature of the boundary conditions between the periods of unperturbed free oscillation. A more complete description of the impact (including, for example, inelastic collisions as in Ben-Reuven[11]) might well account for the discrepancies between data and model in Fig.3. However, we do not know of any way to predict the initial slope of $b(p)$, nor its functional form below and above the quenching, from a specification of the boundary conditions. Neither the mathematics of the noisy oscillator (see, e.g. the book by Gittenberg[13]) nor of the classical kicked rotor (see, e.g. *Ref.*) appear to answer this question.

The key result is that the ammonia model (Fig.3) and even the broken sinusoid of Fig.4(c) both show the Rabi oscillation frequency shifting to lower frequency, broadening, and quenching – going to zero frequency – as the impact rate is increased, in agreement with experiment. Note that the density matrix shows no evidence of localisation. In particular, the off-diagonal elements do not go towards zero as predicted by the decoherence programme. Nor is there any evidence or quantum state diffusion towards the configurational eigenstates. The ammonia shift and quenching are fully accounted for in terms of a perturbed oscillator, and should not therefore be cited as an experimental observation of quantum localisation.

*Acknowledgements:*—We are grateful to Prof. I.C. Percival and Dr T. Prellberg for useful discussions.